\documentstyle[11pt,titlepage]{article}

\begin{document}
\begin{titlepage}
\title{\bf Self-consistent electron subbands of GaAs/AlGaAs
heterostructure in magnetic fields parallel to the interface}

\author{ T. Jungwirth,  L. Smr\v{c}ka \\
{\normalsize Institute of Physics, Acad. of Sci. of Czech Rep.,} \\
{\normalsize Cukrovarnick\'{a} 10, 162 00 Praha  6, Czech Republic}}
\date{\normalsize  J.Phys.: Condens. Matter 1993 (in press) \\
\vspace{6cm}
\noindent{\large PACS numbers: 73.40L, 73.40}}
\maketitle
\end{titlepage}
\begin{titlepage}
\vspace*{5cm}
\noindent {\bf Abstract.}
The   effect  of   strong  magnetic   fields  parallel   to  $\rm
GaAs/Ga_{x}Al_{1-x}As$  interface  on  the  subband  structure of
a 2D  electron layer  is investigated  theoretically. The  system
with two levels occupied in zero magnetic field is considered and
the  magnetic field  induced depletion  of the  second subband is
studied.  The   confining  potential  and   the  electron  energy
dispersion   relations  are   calculated  self-consistently,  the
electron  - electron  interaction is  taken into  account in  the
Hartree approximation.
\end{titlepage}

Recently, there was suggested \cite{1} that the deviations of the
2D  Fermi line  from the  circular shape,  which are  due to  the
combined influence of an  approximately triangular potential well
and of the parallel magnetic field, may play an important role in
the  theory  of  magnetotransport  in  two-dimensional  inversion
layers    at    interfaces    of    $\rm   GaAs/Al_{x}Ga_{1-x}As$
heterostructures.

Zawadzki,  Klahn and  Merkt \cite{2}  studied theoretically Fermi
lines of narrow-band-gap semiconductors  of the In-Sb type. Their
analysis is  based on two main  simplifications: i) The confining
potential  $V_{conf}(z)$ is  taken in  the form  of a  triangular
potential   well.  ii)   The  interface   is  considered   as  an
unpenetrable hard  wall which is accounted  for by an appropriate
boundary  condition  for  the  wave  function.  In this model the
spacing of  levels due to the  magnetic quantization is inversely
proportional  to the  effective mass  $m$, while  the spacing  of
subbands  due  to  the  confining  potential  is  proportional to
$(1/m)^{1/3}$.  Thus,  for  narrow-gap  semiconductors with small
effective  masses  the  electric   and  magnetic  effects  become
comparable for not to high magnetic  fields and for this range of
fields also  the change of Fermi  lines from the circular  to the
\lq egg-like\rq{} form is expected.

In semiconductors with larger effective mass like GaAs the effect
of  parallel  magnetic  field  should  be  less  pronounced  and,
therefore,  it was  till now  treated mainly  by the perturbation
theory \cite{3}. On  the other hand it is well  known that due to
the  small   conduction  band  offset   between  GaAs  and   $\rm
Al_{x}Ga_{1-x}As$ the interface wall is rather soft and electrons
can partly penetrate into $\rm Al_{x}Ga_{1-x}As$. Thus, to obtain
quantitatively better  results, it is desirable  to go beyond the
triangular well  approximation or the  perturbation approach. For
this  reasons   we  decided  to   performe  the  full   numerical
self-consistent study  of the influence of  the in-plane magnetic
field on  the subband shape of  2D electron gas confined  to $\rm
GaAs/Al_{x}Ga_{1-x}As$  interface and  this contribution presents
its results.

It  is now  well-established that  the numerical  self-consistent
calculation based  on the effective  mass approximation describes
correctly  the observed  electron subband  structure in  the zero
magnetic  field.  Moreover,  generally  accepted  methods of such
calculation including  the Hartree approximation  should be valid
in  the presence  of the  magnetic field  as well  as in the zero
field   case.   The   self-consistency   requirement,   i.e.  the
requirement  of  solving   coupled  Poisson  and  Schr\"{o}dinger
equations,  is   an  important  aspect  of   the  energy  spectra
calculations in doped heterostructures as the charge distribution
reacts  on the  confining potential  which itself  determines the
charge distribution. In strong magnetic  fields there is one more
reason for the self-consistency, the charge redistribution caused
by the magnetic field.

For doped $\rm  GaAs/Al_{x}Ga_{1-x}As$ heterostructures the total
charge density $\varrho(z)$ entering  the Poisson equation can be
splitted into parts corresponding to concentrations of electrons,
their  parent  donors  in   $\rm  Al_{x}Ga_{1-x}As$  and  ionized
residual acceptors in GaAs:

\begin{equation}
\varrho(z)=e\left[N_e(z)-N_d^+(z)+N_a^-(z)\right].
\end{equation}

\noindent We  accept a usual  approximation of constant  impurity
concentrations  and assume  donors  and  acceptors to  be ionized
within certain  finite intervals $l_d$  and $l_a$: $N_d^+(z)=N_d$
for $-l_d-w\leq  z\leq -w$ and  $N_a^-(z)=N_a$ for $0\leq  z \leq
l_a$, $w$ is  the spacer thickness (see figure  1). The confining
potential

\begin{equation}
V_{conf}(z)=V_b(z)+V_{s.c.}
\end{equation}

\noindent is  a sum of  the step function  $V_b(z)=V_b\Theta(-z)$
corresponding  to the  conduction band  discontinuity and  of the
Hartree term $V_{s.c.}(z)$ determined from the Poisson equation

\begin{equation}
\frac{d^{2}V_{s.c}}{dz^{2}}=-\frac{\varrho(z)}{\varepsilon} \,\, .
\end{equation}

\noindent  The conduction  band offset  $V_b$ and  the dielectric
constant   $\varepsilon$   enter   our   calculations   as  input
parameters.

The simplest  semiempirical model working  quantitatively for the
lowest   conduction   states   of   $\rm   GaAs/Al_{x}Ga_{1-x}As$
heterostructures is used to solve the Schr\"{o}dinger equation in
the  envelope function  approximation. The  envelope function  is
assumed  to  be  built  from  host  quantum  states  belonging to
a single  parabolic  band.  The  effect  of  the  effective  mass
mismatch is  completely neglected and  the envelope functions  of
GaAs  and  $\rm  Al_{x}Ga_{1-x}As$  are  smoothly  matched at the
interface.

Due to the  translational invariance in the layer  plane the wave
function $\psi_{\alpha}({\bf r})$ can be factorized

\begin{equation}
\psi_{\alpha}({\bf r})=\frac{1}{\sqrt{S}}e^{i(k_xx+k_yy)}
\varphi_{i,k_x}(z)
\end{equation}

\noindent and the Schr\"{o}dinger equation may be written as

\begin{eqnarray}
\left    [-\frac   {\hbar^{2}}    {2m}   \frac   {\partial
^{2}}{\partial z^{2}}+ \frac{1}{2m}\left(\hbar k_{x}-\mid e\mid B_{y}
z\right)^{2} - e V_{conf}\left(z\right)\right]\varphi_{i,k_{x}}
\left(z\right)=\nonumber \\
=\left[ E_{i}(k_x)  -\frac{\hbar^{2}k_{y}^{2}}{2m} \right]
 \varphi_{i,k_{x}}\left(z\right)
\end{eqnarray}

\noindent where  the magnetic field  parallel to the  layer plane
$x-y$   has   a   form   ${\bf   B}\equiv   (0,B_y,0)$.   As  the
heterostructure is  in an electric  and thermodynamic equilibrium
two additional  conditions must be fulfilled:  the charge balance
condition  and  the  constant  chemical  potential condition. For
details about the self-consistent procedure for the zero magnetic
field (but relevant  for $B\neq 0$ as well) we  refer to works by
Stern and Das  Sarma \cite{4} and Ando \cite{5}.  From the latter
paper we also took values of the band offset $V_{b} = 300$meV and
the dielectric constant $\varepsilon =12.9$.

To distinguish  between the charge  redistribution resulting from
the  standard self-consistent  loop  in  zero magnetic  field and
changes   due   to   the   magnetic   field   we   performed  the
self-consistent  procedure  in  two  steps.  First,  the electron
structure  was  calculated  self-consistently  for  $B  =  0$ and
$N_e(z)$ and $V_{conf}$ obtained  from this calculation were used
as input values for the  electron structure determined for $B\neq
0$. In this way the \lq intermediate\rq{} results ($B\neq 0$, but
self-consistency only for  $B = 0$) were obtained.  In the second
step  the   procedure  continued  by   the  full  self-consistent
calculation for $B\neq 0$ yielding the final resuls.

The    parameters    $N_d=2\times     10^{18}    \rm    cm^{-3}$,
$N_a=10^{13}\rm cm^{-3}$  and $w=2\rm nm$  were chosen to  obtain
the electron system of  $N_e\approx 14\times 10^{11} \rm cm^{-2}$
having two  levels occupied in zero  magnetic field and depleting
the second level at $\approx 10\rm T$. Note that in this case the
\lq  intermediate\rq{}  results  are  all  based  on the electron
density  corresponding  to  two  occupied  levels  while the full
self-consistent  study   includes  the  magnetic   field  induced
transfer of  electrons from the  second subband to  the first one
into  the   solution  of  coupled   Poisson  and  Schr\"{o}dinger
equations.

In the zero magnetic field the in-plane electron motion and its
out-of-plane component along the $z$-axis are completely independent.
It means that all electrons
within a subband are described by a single localized wavefunction
$\varphi_{i}\left(z\right)$ regardless their energies $E_{i}(k_{x})$
and  wavevectors  $k_x$, $k_y$. In our case, with two occupied  subbands,
we have two different wavefunctions; their
centres of mass $\langle z\rangle_0$ and $\langle z\rangle_1 $
determine the averaged distances  of electrons in subbands
from the interface.

When    the   magnetic    field   is    applied   the   effective
electro-magnetic potential  $V_{eff}$ composed from  the harmonic
magnetic  potential  and  the  confining  potential $V_{conf}$ is
built

\begin{equation}
V_{eff} = \frac{m \omega^2}{2}\left(z -z_0\right)^{2} - e V_{conf}(z) .
\end{equation}

\noindent  The  centre  $z_{0}$  of  the  magnetic  part  of  the
effective  potential  is  related  to  the  wave vector component
$k_{x}$  by $z_0  = \hbar  k_x/m\omega$. Thus  the magnetic field
couples  the electron  motion in  $x$ and  $z$ directions and for
each   $k_x$   new   $\varphi_{i,k_{x}}\left(z\right)$   must  be
calculated. Also the energy  spectrum $E_{i}(k_{x})$ will deviate
from   the   original   parabolic   dependence   on   $k_x$   and
$E_{i}(k_{x})  \neq E_{i}(-k_{x})$  due to  the breakdown  of the
time  reversal symmetry.  This  results  in the  asymmetric Fermi
lines.  New eigenfunctions  $\varphi_{i,k_{x}}\left(z\right)$ are
shifted  from their  original positions  $\langle z\rangle_0$ and
$\langle z\rangle_1 $  obtained for $B = 0$  and, therefore, also
the charge distribution described by their squares is changed.

There exists  a relation between the  centre of mass of  the wave
function $\varphi_{i,k_{x}}\left(z\right)$  and the shape  of the
energy spectrum curve $E_{i}(k_{x})$

\begin{equation}
\langle z\rangle_{i,k_x} = \frac{\hbar k_{x}}{m\omega} -
\frac{1}{\hbar \omega}\frac{\partial E_{i}(k_{x}) }{\partial k_x}
\end{equation}

\noindent  which makes  it  possible  to calculate  this quantity
without  numerical   difficulties.  As  there   is  one  to   one
correspondence between  $\langle z\rangle_{i,k_x}$ and  $k_x$ and
as we are interested mainly  in the charge redistribution induced
by  the in-plane  magnetic field,  Figure 2  presents the  energy
subbands  as functions  of $\langle  z\rangle_{i,k_x}$ instead of
$k_x$, together  with the shape of  the self-consistent potential
$V_{conf}$. The dashed lines  correspond to \lq intermediate\rq{}
results  while the  full lines  describe the  results of the full
self-consistent calculations.

We already  mentioned that in  the zero magnetic  field electrons
reach  all  possible  energies  for  just  two  centres  of  mass
$\langle  z\rangle_{0}$   and  $\langle  z\rangle_{1}$.   If  the
magnetic field is applied the  electrons with low energies remain
localized approximately around these two values but the electrons
with energies closer to the Fermi level are shifted either to the
interface or deep into the  bulk GaAs, depending on the direction
of  their  motion.  The  minimum  and  maximum distances $\langle
z\rangle_{i,k_x,min}$  and   $\langle  z\rangle_{i,k_x,max}$  are
reached for  the Fermi energy and  correspond to electrons moving
in opposite directions. Note that  the velocity is related to the
energy spectrum by

\begin{equation}
\langle v\rangle_{i,k_x} =
\frac{1}{\hbar}\frac{\partial E_{i}(k_{x}) }{\partial k_x} .
\end{equation}

\noindent The electrons in the  second subband are more sensitive
to the  magnetic field and  the second subband  is emptied at  $B
=11.8$T  for \lq  intermediate\rq{} and   at $B  = 12$T  for full
self-consistent calculation.

The difference  between two types of  calculations increases with
increasing  magnetic  field.  This   is  particularly  valid  for
depressions  of the  confining potential  far from  the interface
which  appear in  results of  full self-consistent  calculations.
Note that  in spite of this  the maximum distance of  an electron
from the interface is not reached for the maximum magnetic field,
but approximately for $B = 12$T.

Figure   3   shows   the   Fermi   lines   based   both   on  \lq
intermediate\rq{} and  full calculations. Two  concentric circles
correspond to two parabolic subbands  in the zero magnetic field.
With  increasing  field  the  area  of  the  second subband Fermi
surface decreases and  the Fermi line of the  first subband takes
the  \lq  egg-like\rq{}  form.  Its  left  half  corresponding to
electrons close to the  interface remains approximately circular,
while  the right  half, describing  the electrons  which move far
from the interface in the  bulk GaAs takes nearly parabolic form.
The gross qualitative features  of both \lq intermediate\rq{} and
full  resuls  are  very  similar.  The  closer  look shows marked
differences  between the  full and  dashed lines.  Because of the
large  magnetic  field  induced  charge  redistribution, the full
lines   are  qualitatively   different  having   rather  the  \lq
pear-like\rq{} than the \lq egg - like\rq{} shapes.

Individual  wave  functions  are  even  more  influenced  by  the
magnetic field.  Figure 4 shows the  first subband wave functions
$\varphi_{0,k_{x}}\left(z\right)$   corresponding   to   $\langle
z\rangle_{0,k_x,min}$  and   $\langle  z\rangle_{0,k_x,max}$.  In
comparison with  the case $B  = 0$ the  function corresponding to
$\langle  z\rangle_{0,k_x,min}$ is  narrower and  shifted to $\rm
Al_{x}Ga_{1-x}As$  spacer and  dotted region,  while the function
corresponding  to  $\langle  z\rangle_{0,k_x,max}$  is  broad and
almost  entirely  inside  the   bulk  GaAs.  Note  a  double-peak
structure of the wave function for $B = 6\rm T$.

We conclude  that as in the  case of the zero  magnetic field the
electron structure of $\rm GaAs/Al_{x}Ga_{1-x}As$ heterostructure
in strong in-plane fields  has to be calculated self-consistently
if  we  are  interested  in  semi-quantitative  results. Both the
energy spectra  and the confining  potential are affected  by the
charge redistribution caused by the  magnetic field. On the other
hand the total number of electrons  remain constant. It is due to
the fact  that the highest  density of states  corresponds to the
bottom part  of the energy  spectra which are  less influenced by
the self-consistency.

\newpage
\thispagestyle{empty}

\newpage
\thispagestyle{empty}
\noindent {\Large{\bf Figure captions}}

\vspace{8mm}
\noindent  {\bf Figure  1.} The  charge distribution  in a single
modulation-doped $\rm  GaAs/Al_{x}Ga_{1-x}As$ heterojunction with
quasi-2DEG at the interface in the depletion length model.

\vspace{5mm}
\noindent   {\bf   Figure   2.}   Electron  eigenenergies  $E(z)$
calculated as  a function of  $\langle z\rangle_{i,k_x}$ and  the
confining potential  $V(z)$. The energies  are measured from  the
Fermi    energy.   Dashed    lines   correspond    to   the   \lq
intermediate\rq{} results, full lines to  the results of the full
self-consistent calculations.

\vspace{5mm}
\noindent  {\bf Figure  3.} Lines  of constant  Fermi energie  in
$(k_x,k_y)$ space are shown  for the two-subband system depleting
the  second  subband  at  $11.8$T  (\lq intermediate\rq{} result)
resp.  $12$T  (full  self-consistent  result).  The  distances in
$k$-space are measured from their centres.

\vspace{5mm}
\noindent  {\bf  Figure  4.}  Wave  functions  of  first  subband
electrons    with   $\langle    z\rangle_{0,k_x,min}$,   $\langle
z\rangle_{0,k_x,max}$ illustrate the charge redistribution due to
in-plane magnetic fields.

\end{document}